\documentclass[10pt,A4paper,amsmath,amssymb,aps,etoolbox,floatfix,longbibliography,
nofootinbib,noshowkeys,noshowpacs,onecolumn,prd,superscriptaddress]{revtex4}

\newcommand{\bea}{\begin{eqnarray}}
\newcommand{\beq}{\begin{equation}}

\newcommand{\eea}{\end{eqnarray}}
\newcommand{\eeq}{\end{equation}}
\newcommand{\ds}{\displaystyle}

\usepackage{float}

\usepackage{tikz}
\newcommand{\dci}[1]{%
    \tikz[baseline=(a.base)]{
        \node[inner sep=0pt] (a) {$#1$};
        \draw (a.north west) ++(0.3em, 0.4ex) circle (0.8pt); 
        \draw (a.north east) ++(-0.2em, 0.4ex) circle (0.8pt);
    }
}
\newcommand{\ci}[1]{\mathring{#1}}

\usepackage{amsmath}
\usepackage{amsfonts}\usepackage{amssymb}
\usepackage{bm}      
\usepackage{comment}
\usepackage{dcolumn}  
\usepackage{epsfig}
\usepackage{graphics}
\usepackage{graphicx} 
\usepackage{lipsum}
\usepackage{natbib}
\usepackage[rightcaption]{sidecap}
\usepackage{placeins}
\usepackage{psfrag}
\usepackage{times}
\usepackage[varg]{txfonts}
\usepackage{xcolor}
\usepackage{xspace}

\graphicspath{%
    {converted_graphics/}
    {/}
}
  
\definecolor{lime}{HTML}{A6CE39}
\DeclareRobustCommand{\orcidicon}{
	\begin{tikzpicture}
	\draw[lime, fill=lime] (0,0) 
	circle [radius=0.16] 
	node[white] {{\fontfamily{qag}\selectfont \tiny ID}};
	\draw[white, fill=white] (-0.0625,0.095) 
	circle [radius=0.007];
	\end{tikzpicture}
	\hspace{-2mm}
}

\begin{document}

\title{Generalised Aichelburg-Sexl and Self-Force for photons}

\author{Abedennour Dib \orcidicon \footnote{ORCID 0000-0003-0089-5062, abedennour.dib@cnrs-orleans.fr}} 

\affiliation{
\mbox{Laboratoire de Physique et Chimie de l'Environnement et de l'Espace (LPC2E) UMR 7328}\\
\mbox{Centre National de la Recherche Scientifique (CNRS), Universit\'e d'Orl\'eans (UO), Centre National d'\'Etudes Spatiales (CNES)}\\
\mbox {3A Avenue de la Recherche Scientifique, 45071 Orl\'eans, France}}

\affiliation{
\mbox{Observatoire des Sciences de l'Univers en region Centre (OSUC) UMS 3116} \\
\mbox{Universit\'e d'Orl\'eans (UO), Centre National de la Recherche Scientifique (CNRS)}
\mbox{Observatoire de Paris (OP), Universit\'e Paris Sciences \& Lettres (PSL)}\\
\mbox{1A rue de la F\'{e}rollerie, 45071 Orl\'{e}ans, France}}

\affiliation{
\mbox{D\'epartement de Physique, UFR Sciences et Techniques,  Universit\'e d'Orl\'eans, Rue de Chartres, 45100 Orl\'{e}ans, France}}

\author{Aymeric Garnier\orcidicon\footnote{ORCID 0009-0000-0855-8875, aymeric.garnier@etu.univ-orleans.fr}}

\affiliation{
\mbox{Laboratoire de Physique et Chimie de l'Environnement et de l'Espace (LPC2E) UMR 7328}\\
\mbox{Centre National de la Recherche Scientifique (CNRS), Universit\'e d'Orl\'eans (UO), Centre National d'\'Etudes Spatiales (CNES)}\\
\mbox {3A Avenue de la Recherche Scientifique, 45071 Orl\'eans, France}}

\affiliation{
\mbox{Observatoire des Sciences de l'Univers en region Centre (OSUC) UMS 3116} \\
\mbox{Universit\'e d'Orl\'eans (UO), Centre National de la Recherche Scientifique (CNRS)}
\mbox{Observatoire de Paris (OP), Universit\'e Paris Sciences \& Lettres (PSL)}\\
\mbox{1A rue de la F\'{e}rollerie, 45071 Orl\'{e}ans, France}}

\affiliation{
\mbox{D\'epartement de Physique, UFR Sciences et Techniques,  Universit\'e d'Orl\'eans, Rue de Chartres, 45100 Orl\'{e}ans, France}}

\affiliation{
\mbox{Scuola Superiore Meridionale,  Universit\`{a} degli Studi di Napoli, Federico II}\\
\mbox{Largo San Marcellino 10, 80138 Napoli, Italy}}

\affiliation{
\mbox{Dipartimento di Fisica E. Pancini, Universit\`{a} degli Studi di Napoli, Federico II}\\
\mbox{Complesso Universitario Monte S. Angelo, Via Cinthia 9 Edificio G, 80126 Napoli, Italy}}

\author{Alessandro D.A.M. Spallicci\orcidicon\footnote{ORCID 0000-0002-8920-4610, spallicci@cnrs-orleans.fr}} 

\affiliation{
\mbox{Laboratoire de Physique et Chimie de l'Environnement et de l'Espace (LPC2E) UMR 7328}\\
\mbox{Centre National de la Recherche Scientifique (CNRS), Universit\'e d'Orl\'eans (UO), Centre National d'\'Etudes Spatiales (CNES)}\\
\mbox {3A Avenue de la Recherche Scientifique, 45071 Orl\'eans, France}}

\affiliation{
\mbox{Observatoire des Sciences de l'Univers en region Centre (OSUC) UMS 3116} \\
\mbox{Universit\'e d'Orl\'eans (UO), Centre National de la Recherche Scientifique (CNRS)}
\mbox{Observatoire de Paris (OP), Universit\'e Paris Sciences \& Lettres (PSL)}\\
\mbox{1A rue de la F\'{e}rollerie, 45071 Orl\'{e}ans, France}}

\affiliation{
\mbox{D\'epartement de Physique, UFR Sciences et Techniques,  Universit\'e d'Orl\'eans, Rue de Chartres, 45100 Orl\'{e}ans, France}}

\date{26 June 2025}

\begin{abstract}
 We generalise the Aichelburg-Sexl Stress-Energy-Momentum tensor for massless particles to include motion with angular velocity,  introducing tensorial spherical harmonics. Further, we make steps towards the concept of self-force for photons which could manifest as frequency shift.
\end{abstract}


\maketitle

\section{Introduction}\label{sec:introduction}

Interactions involving massless particles occur in astrophysical processes, black hole and gravitational wave phenomena. Their analysis may induce questioning fundamental issues in physics. 

We consider the pioneer work by Aichelburg and Sexl \cite{Aichelburg-Sexl-1971}, closely followed by Dray and 't Hooft \cite{Dray-tHooft-1985} (for some more recent applications see e.g. \cite{Esposito-Pettorino-Scudellaro-2007,Battista-Esposito-Scudellaro-Tramontano-2016}. They describe the gravitational field of an ultra-relativistic massless particle boosted to the speed of light. The result has been obtained in two concurring ways. The former consists in solving the linearised Einstein field equations for a particle with a rest mass $m$ tending to zero and moving uniformly with the velocity $v$ tending to $c$, thereby keeping finite the energy of the particle. This is equivalent to solving the field equations directly with a Stress-Energy-Momentum (SEM) tensor for a massless particle as source term.
The latter acts in the Schwarzschild-Droste (SD) metric \cite{Schwarzschild-1916,Droste-1917}\footnote{The SD metric was obtained independently by the two authors in the same year \cite{Rothman-2002}.}. Aichelburg and Sexl apply to the SD metric a Lorentz transformation to obtain the gravitational
field as seen by an observer moving uniformly relative to the mass. Again the limits $v\rightarrow c$ and $m \rightarrow 0$ are imposed. The authors show that the linearised solution and the exact solution are in agreement. 

However, in its original formulation, Aichelberg and Sexl work did not account for angular motion, which is crucial in many astrophysical and theoretical contexts, such as particle collisions with non-zero impact parameters and ultra-relativistic scattering in curved spacetime. Further, they did not make use of tensorial spherical harmonics (TSH). Thus, in this work, we generalise the Aichelburg-Sexl formalism for massless particles by incorporating angular velocity into the SEM tensor, and adopting the source term in the Regge-Wheeler-Zerilli (RWZ) equations. The RWZ equations describes non-rotating black hole perturbations in SD spacetime. The first study in absence of a source \cite{Regge-Wheeler-1957} verified the return to stability after a kick to the back hole; later in presence of a radially falling particle \cite{Zerilli-1970a,Zerilli-1970b,Zerilli-1970c}, the radiated energy and the spectrum were obtained through an analysis in Fourier domain; finally, non-zero angular momentum was considered during plunging \cite{Detweiler-Szedenits-1979}. The RWZ equations have been applied and reworked in all directions from the early days: for gravitational waveforms and radiated energy, see the so-called DRuPP \cite{Davis-Ruffini-Press-Price-1971}, while for gauge invariance, see \cite{Moncrief-1974}. 

By extending the description in \cite{Aichelburg-Sexl-1971} to accommodate massless particles with arbitrary motion, we provide a more comprehensive framework for studying gravitational interactions.

We revisit the Aichelburg-Sexl metric and its derivation and construct a modified SEM tensor that accounts for angular motion. Using this new tensor, we derive the corresponding source terms in the RWZ formalism and present a detailed analysis of the resulting perturbation equations. Our approach follows the established method of decomposing perturbations into TSH, allowing us to systematically examine both even- and odd-parity modes \cite{Regge-Wheeler-1957,Zerilli-1970a,Zerilli-1970b,Zerilli-1970c}.

This paper has a second objective, that is the the description of how curvature affects the motion and the energy of a massless particle as a photon, that is the self-force. Thereby, we investigate the self-force of a massless particle in black hole geometry. Furthermore, there is a never ending interest for the gravity of light, {\it e.g.}, \cite{Bonnor-1969,vanHolten-2018}.   

Let us recall the conceptual steps that lead to determine the motion of a particle subjected to the self-force \cite{Spallicci-Ritter-Aoudia-2014}. At lowest order, a massive particle follows the geodesic of the background metric. At this order in motion, the system (particle plus the main body) emits gravitational radiation. At higher order, the latter impinges back on the motion of the particle, proportionally to the mass ratio (particle/main body); please note that the particle is already affected by the non-radiative mass ratio correction of Newtonian origin \cite{Spallicci-vanPutten-2016} corresponding to the $\ell = 0, 1$ modes in general relativity. 
These corrections act as a force term that displaces the particle from the background geodesic motion and are quantitatively described by the MiSaTaQuWa equation \cite{Mino-Sasaki-Tanaka-1997,quwa97}. An alternative point of view came with the work of Detweiler and Whiting \cite{Detweiler-2001,Detweiler-Whiting-2003}, who considered the particle moving geodesically in the full metric (background plus perturbations). They further demonstrated that the metric perturbation can be decomposed into singular and regular parts, with the latter governing the effective geodesic motion. 

In this brief account, we ought to mention the mode-sum regularisation method \cite{Barack-Ori-2000} to deal with divergencies arising from the point-like nature of the falling particle; the benefits arising from the numerical time-domain simulation method \cite{Lousto-Price-1997a,Lousto-Price-1997b} and the implementation of the iterative self-consistent method for corrections in radial fall \cite{Ritter-Aoudia-Spallicci-Cordier-2016b} proposed by Gralla and Wald \cite{Gralla-Wald-2008}. Curiously, it took exactly a century to properly compute the effect of the self-force in the extremely non-adiabatic radial fall case.  

From the Orl\'eans Capra conference in 2008 \cite{Blanchet-Spallicci-Whiting-2011}, the issue of the self-force has been studied comparing perturbation methods, Effective One-Body and Parametrised post-Newtonian methods; lately for massless particles \cite{Capozziello-Menedeo-Usseglio-2025}. 

The paper is structured as follows: Section \ref{sec:TSH-RWZ} reviews the established framework of tensorial spherical harmonic decomposition including the derivation of the RWZ equations. In Section \ref{sec:massless} we present our core contributions: \ref{sec:ASmodif} the generalisation of the Aichelburg-Sexl SEM tensor to include angular velocity for massless particles, \ref{sec:RWZmassless} the derivation of the modified RWZ equations with angular-motion source terms. In Section \ref{sec:SF}, we describe the concept of self-force and express how the self-force may act on a massless particle. Finally, Section \ref{sec:Conclusion} concludes the work.

\section{Tensorial Spherical Harmonic decomposition~and the Regge-Wheeler-Zerilli equations}\label{sec:TSH-RWZ}

The SD geometry is best suited for the utilisation of TSH when solving the linearised Einstein equations. The SEM tensor and the ten metric perturbation $h_{\alpha\beta}$ are decomposed into TSH on a multipolar basis \cite{Regge-Wheeler-1957,Zerilli-1970a,Zerilli-1970b,Zerilli-1970c} and \cite{Martel-2004,sanasa06}. The SEM is given by  

\begin{equation}
    T_{\alpha\beta}=\sum_{\ell m}\sum_i^{10} T^{(i)\ell m}(t,r)Y_{\alpha\beta}^{(i)\ell m}(\theta,\phi)~~,
    \label{eq:Tdecomp}
\end{equation}
where 
\begin{align}
T^{(i)\ell m}(t,r)=\int_{\mathcal{S}^2}g^{\alpha \delta}g^{\beta\gamma}T_{\alpha\beta}\left(Y_{\delta\gamma}^{(i)\ell m}\right)^*d\Omega~~,
\end{align}
and $d\Omega = \sin\theta d\theta d\phi$; the asterisk indicates complex conjugation. The spherical harmonics are defined as
\begin{align}
Y_{\ell m}(\theta, \phi)=\sqrt{\frac{2\ell +1}{4\pi}\frac{\ell -m}{\ell +m}}P_{\ell m} (\cos\theta)e^{im\phi}=N_{\ell m}P_{\ell m} (\cos\theta)e^{im\phi}~~,
\end{align}
$P_{\ell m} (\cos\theta)$ being the associated Legendre polynomials defined in the range:
$\theta \in [0,\pi],$ $\phi \in [0,2\pi)$. The same decomposition applies to the perturbation tensor $h_{\alpha\beta}$

\begin{equation}
    h_{\alpha\beta} = \sum_{\ell m} \sum_{i=1}^{10} h^{(i)\ell m}(t,r) Y^{(i)\ell m}_{\alpha\beta}(\theta,\phi)~~.
    \label{eq:hdecomp}
\end{equation}

The functions $h^{(i)\ell m}(t,r)$ are split into even $(-1)^\ell$ and odd parity $(-1)^{\ell+1}$ modes. We thus write 

\begin{equation}
h_{\alpha\beta} = \sum_{\ell m}\left[h^{(o)\ell m}_{\alpha\beta} + h^{(e)\ell m}_{\alpha\beta}\right]~~.
\label{eq:hparitydecomp}
\end{equation}

When we insert the polar decomposition into the linearised Einstein equations, we obtain a system of equations that can be reduced to a single master equation for each parity sector (incidentally, this occurs in the RW gauge and not in the Lorenz gauge where the self-force was originally conceived; nevertheless, there are correspondences to pass from one gauge to the other).
For even parity, we have seven equations with four unknowns, while for odd parity, we have three equations with two unknowns. 

The RWZ master for each parity type is defined as

\begin{equation}\label{eq:RWZ}
\left[\frac{\partial^2}{c^2\partial t^2} - \frac{\partial^2}{\partial r^{*\,2}} + V_{\ell}^{(e/o)}(r)\right]\psi_{\ell m}^{(e/o)}(r,t)= - S_{\ell m}^{(e/o)}(r,t), \quad r_*=r+\frac{2GM}{c^2}\ln\left(\frac{rc^2}{2GM}-1\right)~~,
\end{equation}
where $r_*$, the tortoise coordinate \cite{Droste-1917}, displaces the black hole apparent horizon singularity to minus infinity. $V_{\ell}^{(e/o)}(r)$ are the Regge-Wheeler (odd) and Zerilli (even) $V (\ell, r, M)$  potentials defined by the background metric and $S_{\ell m}^{(e/o)}(r,t)$ are the source terms derived from the SEM tensor 

\beq
S_{\ell m}^{(e/o)}(r,t)= G(r,t)\delta[r-r_{p}(t)]+F(r,t)\delta '[r-r_p(t)]~~,
\eeq
where the prime denotes an $r$ derivative, $r_p(t)$ the radial position of the particle as a function of time, and
$G$ and $F$ are functions of $r$ and $t$ depending on the
particle trajectory. In the jargon, the $\psi$ wave-function belongs to the $C^{-1}$ continuity class, due to the $C^{-3}$ presence of the Dirac distribution derivative. This complexity obliges to deal numerically with the RWZ equations in time domain. The transformation into Fourier domain eases the finding of a solution, but leaving out low or high frequencies for a semi-analytical treatment may lead to undetermined errors in the trajectory when returning to the time domain.  

Finally, the wave-functions $\psi_{\ell m}^{(e/o)}(r,t)$ to this equation are used to reconstruct the metric perturbations. 

There are different definitions and formulations of TSH and often plagued by computational errors. We feel confident in choosing the prescription by Martel \cite{Martel-2004}. 

\section{The massless particle}\label{sec:massless}

\subsection{Modification of Aichelburg and Sexl SEM tensor}\label{sec:ASmodif}
Aichelburg and Sexl \cite{Aichelburg-Sexl-1971} found that the gravitational field of a rapidly moving particle is dilated in
the direction orthogonal to the particle motion and compressed in the
direction of the motion. The same conclusion was achieved by Penrose and Bonnor \cite{Penrose-1968,Bonnor-1969}. In the
limit of a massless point particle moving with the speed of light, this
compression becomes extreme and the pulsed field is non-vanishing only on a
plane containing the particle, that is a null hypersurface orthogonal to the trajectory, exhibiting a delta-function singularity. The solution belongs to the class of pp-wave metrics and is of Petrov type N, characteristic of plane-fronted gravitational waves \cite{Petrov-2000,Pirani-1959}\footnote{In general relativity, the Ricci tensor is related to the presence of matter; in the absence of matter, the Ricci tensor is zero. But the Weyl tensor is composed by the sum of the Riemann tensor plus contributions depending on the Ricci tensor and scalar. This property gives the Weyl tensor its importance: its structure gives the entire structure of the gravitational field in regions devoid of matter. For example, a region of space crossed by a gravitational wave has a non-zero Weyl tensor. Physically, the Petrov classification is a way to characterise a spacetime by the number of principal null directions it admits. Mathematically, it is the description of the possible algebraic symmetries of the Weyl tensor at each event in a Lorentzian manifold. In the N type, there is a single principal null direction of multiplicity 4.}. 

The SEM tensor of a point particle of rest mass $m$ moving with constant velocity $v$ along the $x$-direction is given by

\beq
    T^{\mu\nu}= \gamma m \delta(x-vt)\delta(y)\delta(z)s^\mu s^\nu~~,
\eeq

where $s^\mu = c\delta_0^\mu + v\delta_1^\mu$ is equivalent to 4-velocity and $\gamma = \left(1 - {\ds \frac{v^2}{c^2}}\right)^{-1/2}$ is the relativistic dilation factor.\\

As $v \to c$, the relativistic factor $\gamma$ diverges. For ensuring a finite SEM tensor, we redefine the mass parameter in terms of the total longitudinal momentum $p$ as

\beq
    m = \frac{p}{\gamma}~~,
\eeq
where $p$ remains finite as $v \to c $. This leads to the ultra-relativistic SEM tensor, {\it i.e.} the Aichelburg-Sexl SEM tensor

\beq
    T^{\mu\nu} = p\delta(x-t)\delta(y)\delta(z)s^\mu s^\nu~~.
\eeq

The Aichelburg-Sexl SEM tensor is well-constructed and useful for describing massless particles. By redefining the mass parameter in terms of the finite longitudinal momentum $p$, the tensor remains finite and well-defined in the ultra-relativistic limit. The delta distributions in the SEM  tensor ensure that the energy is concentrated on a null hypersurface, consistent with the physical picture of a massless particle moving at the speed of light. 

We want to generalise the SEM tensor of Aichelburg-Sexl to include motion with angular velocity. Due to the symmetries of the system and the spherical harmonic decomposition, we adopt spherical coordinates. The new SEM tensor now reads

\beq\label{eq:STAS}
    T^{\mu\nu}=\frac{p}{c^2r^2\sin\theta}\delta(ct-r)\delta(\theta-\theta_p)\delta(\phi-\phi_p)s^\mu s^\nu~~,
\eeq
with $s^\mu = c\delta_0^\mu + c\delta_1^\mu + r\dot{\theta}\delta_2^\mu + r\sin\theta\dot{\phi}\delta_3^\mu~~$.

The angular terms permit the inclusion of the angular velocity which plays a decisive role in most astrophysical processes. 

\subsection{Regge-Wheeler-Zerilli like equations for massless particle}\label{sec:RWZmassless}

Aichelburg and Sexl \cite{Aichelburg-Sexl-1971} did not use TSH, as their analysis did not concern black hole perturbations. This is a first feature, that we pursue adopting the conventions contained in \cite{Martel-2004}; b) a second difference with respect to \cite{Aichelburg-Sexl-1971} is the inclusion of angular velocity; c) finally, the third difference with \cite{Martel-2004} is that our work refers to a massless particle. We will make use of the following definitions coming from the TSH decomposition  
\[
f(r)=1-\frac{2GM}{c^2r}~~,
\]
\[ 
Q^{ab} = 8\pi \int_{S^2} T^{ab} Y^{\ell m*}\, d\Omega~~, \qquad
Q^a = \frac{16\pi r^2}{l(l+1)}\, \int_{S^2} T^{aA} Z^{\ell m*}_A\, d\Omega~~,
\]
\[
Q^\flat = 8\pi r^2 \int_{S^2} T^{AB} U^{\ell m*}_{AB}\, d\Omega~~, \qquad
Q^\sharp = \frac{32\pi r^4}{(l-1)l(l+1)(l+2)}\, \int_{S^2} T^{AB}
V^{\ell m*}_{AB}\, d\Omega~~, 
\]  
\[ 
P^a = \frac{16\pi r^2}{l(l+1)}\, \int_{S^2} T^{aA} X^{\ell m*}_{A}\, d\Omega~~,
\qquad  
P = \frac{16 \pi r^4}{(l-1)l(l+1)(l+2)}\, \int_{S^2} T^{AB} W^{\ell m*}_{AB}\,
d\Omega~~, 
\]
\[
Z^{\ell m}_{A} \ =\ Y^{\ell m}_{|A}~~, \quad X^{\ell m}_{A}\ =\ \varepsilon_{A}^{\ B}Y^{\ell m}_{|B}~~, \quad 
U^{\ell m}_{AB}\ =\ \Omega_{AB}Y^{\ell m}~~, \quad V^{\ell m}_{AB}\ = \ Y^{\ell m}_{|AB}+\frac{l(l+1)}{2}\Omega_{AB}Y^{\ell m}~~, \textrm{  and  } W^{\ell m}_{AB}\ = \ X^{\ell m}_{(A|B)}~~.
\]
Lower-case indices run over $t$ and $r$ while the capital roman indices run over the angular coordinates, the integration is over the unit 
two-sphere with $d\Omega = \sin\theta\, d\theta d\phi$, and $T^{ab}$, $T^{aA}$ and $T^{AB}$ are components 
of $T^{\mu \nu}$. The $|$ denotes a covariant derivative compatible with $\Omega_{AB}$, and $\varepsilon_{AB}$ is the Levi-Civita tensor on $S^2$. The $P$ and $Q$ terms used in this section are the source terms arising from the decomposition of the SEM tensor into spherical harmonics: the $Q$ terms refer to the even-parity components, and the $P$ terms to the odd-parity components. From  Eq. (\ref{eq:STAS}), for $s^0=s^1$, the following identities occur

\begin{align}
    T^{tt}=T^{rr}=T^{rt}=T^{tr}=T^{rr},
\end{align} 
leading to the same equality for the Q coefficient
\begin{align}
    Q^{tt}=Q^{rr}=Q^{rt}=Q^{tr}=Q^{rr}~and ~Q^t=Q^r~.
\end{align} 

We note $\delta(\theta-\theta_p)\delta(\phi-\phi_p)$ by $\delta^2_{\theta,\phi}$, where the subscript $p$ refers to the particle position.

\begin{align}
Q^{tt}=8\pi\int_{S^2}\frac{p}{c^2r^2\sin\theta}\delta(ct-r)\delta^2_{\theta,\phi}Y^{\ell m*}\sin\theta d\theta d\phi=\frac{8\pi p}{r^2}Y^{\ell m*}_p\delta(ct-r)~~.
\end{align}

For $Q^t$ and $Q^r$, we need to compute $Z_\phi^{\ell m}$ and $Z_\theta^{\ell m}$ . Since $Y^{\ell m}$ is a scalar, its covariant derivative is equivalent to its partial derivative so we have:

\begin{align}
Z_\theta^{\ell m}=Y^{\ell m}_{|\theta}=\frac{\partial}{\partial\theta}N^{\ell m}P^{\ell m} (\cos\theta)e^{im\phi}=\frac{\partial\cos\theta}{\partial\theta}\frac{\partial P^{\ell m} (\cos\theta) }{\partial\cos\theta}=-N^{\ell m} \sin\theta \ci{P}^{\ell m}e^{im\phi}~~,
\end{align}

\begin{align}
Z_\phi^{\ell m}=Y^{\ell m}_{|\phi}=imY^{\ell m}~.
\end{align}

The notation $\ci{}$ indicate a derivation with respect to $\cos{\theta}$ leading to

\begin{align}
Q^t=Q^r&=\frac{16\pi r^2}{\ell(\ell+1)}\int_{S^2}\left(T^{t\theta}Z_\theta^{\ell m*}+T^{t\phi}Z_\phi^{\ell m*}\right)\sin\theta d\theta d\phi  \nonumber \\
&=\frac{-16\pi p}{c^2\ell(\ell+1)}\delta(ct-r)\left[(c+r\dot{\theta})^2N^{lm}\sin\theta_p\ci{P}^{\ell m}_pe^{-im\phi_p}+(c+r\sin\theta_p\dot{\phi})imY^{\ell m*}_p\right]~.
\end{align}

The computation of $Q^\flat$  follow the same path

\begin{align}
&Q^\flat=8\pi r^2\int_{S^2}\left[\frac{p}{c^2r^2\sin\theta}\delta(ct-r)\delta^2_{\theta,\phi}r^2\dot{\theta}^2U^{\ell m*}_{\theta\theta}+\frac{p}{c^2r^2\sin\theta}\delta(ct-r)\delta^2_{\theta,\phi}r^2\sin\theta^2\dot{\phi}^2U^{\ell m*}_{\phi\phi}\right]\sin\theta d\theta d\phi~~,\nonumber\\
&Q^\flat=\frac{8\pi r^2 p}{c^2}\delta(ct-r)\left[\dot{\theta}^2Y^{\ell m*}_p+\sin\theta^4\dot{\phi}^2Y^{\ell m*}_p\right]~.
\end{align}

The angular velocity components $\dot{\theta}$ and $\dot{\phi}$ appear explicitly, demonstrating how angular motion directly affects the source term. For $Q^\sharp$, we have to compute $V_{AB}^{\ell m}$ by looking at the second derivative of the spherical harmonics

\begin{align}
Y^{\ell m}_{|\theta\theta}=N^{\ell m}\left(\sin\theta^2\dci{P}^{\ell m}-\cos\theta \ci{P}^{\ell m}\right)e^{im\phi}~~,
\end{align}
\begin{align}
Y^{\ell m}_{|\phi\phi}=-m^2Y^{\ell m}, \quad Y^{\ell m}_{|\theta\phi}=Y^{\ell m}_{|\phi\theta}=-imN^{\ell m}\sin\theta\ci{P}^{\ell m}e^{im\phi}~~.
\end{align}

The computation of the coefficient is straightforward 
\begin{align}
Q^\sharp=\frac{32\pi r^4}{(\ell-1)(\ell+1)(\ell+2)}\int_{S^2}\left[T^{\theta\theta}V_{\theta\theta}^{\ell m*}+2T^{\theta\phi}V_{\theta\phi}^{\ell m*}+T^{\phi\phi}V_{\phi\phi}^{\ell m*}\right]d\Omega~~,
\end{align}

leading to

\begin{align}
Q^\sharp=\frac{32\pi r^4 p\delta(ct-r)}{c^2(\ell-1)(\ell+1)(\ell+2)}
\left[ 
\dot{\theta}^2\left(Y^{\ell m*}_{|\theta\theta p}+\frac{\ell(\ell+1)}{2}Y^{\ell m*}_p\right)
+ 2\dot{\theta}\dot{\phi}\sin\theta_p Y^{\ell m*}_{|\theta\phi p}
+ \sin^2\theta_p\dot{\phi}^2\left(\frac{\ell(\ell+1)}{2}\sin\theta_p Y^{\ell m*}_p -m^2Y^{\ell m*}_p\right)
\right]~~,
\end{align}

\begin{align}
Q^\sharp=\frac{32\pi r^4 p}{c^2(\ell-1)(\ell+1)(\ell+2)}\delta(ct-r)O^{\ell m}_{\theta,\phi}~~.
\end{align}

With:

\begin{align}
O^{\ell m}_{\theta,\phi}=\left[ 
\dot{\theta}^2\left(Y^{\ell m*}_{|\theta\theta p}+\frac{\ell(\ell+1)}{2}Y^{\ell m*}_p\right)
+ 2\dot{\theta}\dot{\phi}\sin\theta_p Y^{\ell m*}_{|\theta\phi p}
+ \sin^2\theta_p\dot{\phi}^2\left(\frac{\ell(\ell+1)}{2}\sin\theta_p Y^{\ell m*}_p -m^2Y^{\ell m*}_p\right)
\right]~~.
\end{align}

This term captures the complex angular dynamics of the massless particle, including cross-terms between $\dot{\theta}$ and $\dot{\phi}$ that represent coupling between polar and azimuthal motion. The computation of the $P$ coefficient is also straightforward. For this we need the Levi-Civita tensor on $S^2$ which is defined as:

\begin{align}
\varepsilon_{AB} =
\begin{pmatrix}
0 & \sin\theta \\
-\sin\theta & 0
\end{pmatrix}~~,
\qquad \varepsilon_{A}^{\ B}=\Omega^{BC}\varepsilon_{AC}=\begin{pmatrix}
    0 & \frac{1}{\sin\theta}\\
    -\sin\theta & 0
\end{pmatrix}~~.
\end{align}

This allows the determination of $P^t$ which is equal to $P^r$ because of the equality of $s^0 = s^1$ defined in Eq. (\ref{eq:STAS})

\begin{align}
X_\theta^{\ell m}=\varepsilon_\theta^{\ \phi} Y^{\ell m}_{|\phi}, \qquad X_\phi^{\ell m}=\varepsilon_\phi^{\ \theta} Y^{\ell m}_{|\theta}~,
\end{align}

\begin{align}
P^t=\frac{-32\pi r p}{c\ell(\ell+1)}\delta(ct-r)\left[\frac{im\dot{\theta}}{\sin\theta_p}Y^{\ell m *}_p+\sin\theta_p^2Y^{\ell m*}_{|\theta p}\right]~.
\end{align}

Finally, for the last coefficient $P$ we need all the $W_{AB}$ 

\begin{align}
W_{AB}=X_{(A|B)}=\frac{1}{2}(X_{B|A}+X_{A|B}), \quad X_{\theta|\theta}=-im\left(N^{\ell m}\ci{P}^{\ell m}e^{im\phi}+\frac{\cos\theta}{\sin\theta^2}Y^{\ell m}\right)~~,
\end{align}

\begin{align}
X_{\phi|\phi}=imN^{\ell m}\sin\theta^2\ci{P}^{\ell m}e^{im\phi}, \quad X_{\phi|\theta}=2N^{\ell m}\sin\theta\cos\theta\ci{P}^{\ell m}e^{im\phi}-\sin\theta^3N^{\ell m}\dci{P}^{\ell m}e^{im\phi}, \quad X_{\theta|\phi}=\frac{-m^2}{\sin\theta}Y^{\ell m}~.
\end{align}

With all these terms we can compute the last coefficient

\begin{align}
P=\frac{16\pi r^4}{(\ell-1)(\ell+1)(\ell+2)}\int_{S^2}\left[T^{\theta\theta}W_{\theta\theta}^{\ell m*} +2T^{\theta\phi}W_{\theta\phi}^{\ell m*} + T^{\phi\phi}W_{\phi\phi}^{\ell m*}\right]d\Omega~.
\end{align}

Once again we factorise the angular part in a single term for simplicity

\begin{align}
P=\frac{16\pi p r^4}{c^2(\ell-1)(\ell+1)(\ell+2)}\delta(ct-r) M_{\theta,\phi}^{\ell m}~~,
\end{align}

such that

\begin{align}
    M_{\theta,\phi}^{\ell m} =& \Bigg[ im\dot{\theta} \left(N^{\ell m}\ci{P}^{\ell m}_p e^{-im\phi_p}+\frac{\cos\theta_p}{\sin^2\theta_p}Y^{\ell m*}_p\right)+ \sin\theta_p\dot{\phi}\dot{\theta} \left(\frac{-m^2}{\sin\theta_p}Y^{\ell m*}_p + 2N^{\ell m}\sin\theta_p\cos\theta_p\ci{P}^{\ell m}_p e^{-im\phi_p} - \sin^3\theta_p N^{\ell m} \dci{P}^{\ell m}_p e^{-im\phi_p} \right)  \notag \\
& -im \sin^4\theta_p\dot{\phi}^2 N^{\ell m} \ci{P}^{\ell m}_p e^{-im\phi_p} \Bigg]~.
\end{align}

Again, by computing the source term, it appears that the $\delta (ct-r) $ and $\delta'(ct-r) $ stand as common term for the same previous reasons; the integration is made only on angular coordinate so only the delta distribution on the null coordinate is left. We can again write the source function as

\begin{equation}\label{eqn:src2}
S^{\ell m}_{(e/o)}=\mathcal{G}_{(e/o)}\delta(ct-r)+\mathcal{F}_{(e/o)}\delta'(ct-r)~.
\end{equation} 

For the $\mathcal{G}$ coefficients we get

\begin{align}
\mathcal{G}_{(e)} = & \frac{1}{(\lambda+1)\Lambda} \Bigg\{ 
r^2 f (f-1) - \frac{16\pi p}{r^3} Y^{\ell m*}_p + (\Lambda - f) \frac{8\pi p}{r} Y^{\ell m*}_p  + \frac{8\pi r^3 f^2 p}{c^2} 
\left[ \dot{\theta}^2 Y^{\ell m*}_p + \sin^4\theta_p \dot{\phi}^2 Y^{\ell m*}_p \right] \notag \\
& - \frac{8\pi p f^2}{r^3 \Lambda} Y^{\ell m*}(\theta_p, \phi_p) 
\left[ \lambda(\lambda-1)r^2 + (4\lambda-9)Mr + 15M^2 \right] 
\Bigg\} \notag \\
& - \frac{32\pi p f}{c^2\Lambda\ell(\ell+1)} 
\Bigg[ (c + r\dot{\theta})^2 N^{\ell m} \sin\theta_p P^{\ell m}_p e^{-im\phi_p} + (c+r\sin\theta_p\dot{\phi}) im Y^{\ell m*}_p \Bigg] - \frac{32\pi r^3 pf}{c^2(\ell-1)(\ell+1)(\ell+2)} O^{\ell m}_{\theta,\phi}~~,\\
\mathcal{G}_{(o)} = & \frac{2f}{r^2} \left( 1 - \frac{3GM}{c^2r} \right) 
\frac{16\pi p r^4 M_{\theta,\phi}^{\ell m}}{c^2(\ell-1)(\ell+1)(\ell+2)}  
 - \frac{32\pi p r^3 f^2M_{\theta,\phi}^{\ell m}}{c^2(\ell-1)(\ell+1)(\ell+2)} -  \frac{32\pi f p}{c\ell(\ell+1)} 
\left[ \frac{im\dot{\theta}}{\sin\theta_p} Y^{\ell m *}_p 
+ \sin^2\theta_p Y^{\ell m*}_{|\theta p} \right]~~,
& 
\end{align}

and for the $\mathcal{F}$ coefficients we get:

\begin{eqnarray}
    &\mathcal{F}_{(e)} =& \frac{8\pi p Y^{\ell m*}}{\Lambda r^2(\lambda+1)} f(f^2-1)~~,\\
    &\mathcal{F}_{(o)}=&\frac{-16\pi p r^3 f^2 M_{\theta,\phi}^{\ell m}}{c^2(\ell-1)(\ell+1)(\ell+2)}~~.
\end{eqnarray}

The above coefficients differ from those discussed in the test particle case. Even though the Aichelburg-Sexl metric is the limit where $v\rightarrow c$ and $m\rightarrow 0$ for a test particle, taking these limits in the expressions derived by Martel for the TSH, will necessarily lead to divergences that cannot be mended. 

\section{The self-force}\label{sec:SF}

In this section, we briefly describe the concept of self-force \cite{Spallicci-Ritter-Aoudia-2014}. 
When a point electric charge moves around another charge of opposite sign, it emits radiation and undergoes radiation reaction spiralling down to the other charge. The radiation propagating outward and the charge spiralling inward represent the retarded soluton and break the time-reversal invariance present in the theory of Maxwell. The advanced solution, produces instead 
a time-reversed picture in which the radiation is propagating inward and the charge is spiralling
outward. Both solutions possess a singularity at the charge.

On the steps of Dirac's work \cite{Dirac-1938}, Detweiler and Whiting \cite{Detweiler-Whiting-2003,Poisson-2011, Poisson-Pound-Vega-2011} have reworked the singular term (from the perturbation field) 
$S\!ing $. This is the mean of the advanced and retarded solutions and it is time-reversal invariant, {\it i.e.} incoming and outgoing energy are equal.  In flat spacetime, the radiative term is obtained by subtracting the singular from the retarded term.  The subtraction deletes the singularity at the particle. The isotropy of the singularity impedes forces actin on the particle. The only remaining term acting on the particle is given by

\begin{align}
Rad = Ret - S\!ing = Ret - \frac{1}{2}[Ret + Adv] = \frac{1}{2}[Ret - Adv]~~.
\end{align}      

\begin{figure}[!ht]
\centering
\begin{minipage}[c]{0.45\textwidth}
  \includegraphics[width=6cm]{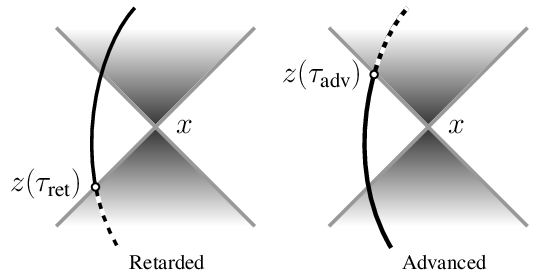}
  \caption{Retarded and advanced terms (dotted line). }
  \label{fig1}
\end{minipage}
\quad
\begin{minipage}[c]{0.45\textwidth}
  \includegraphics[width=2.6cm]{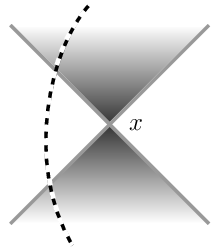}
  \caption{The H term (dotted line). }
  \label{fig2}
\end{minipage}

\end{figure}

\begin{figure}[!ht]
  \centering
\begin{minipage}[c]{0.45\textwidth}
  \includegraphics[width=3cm]{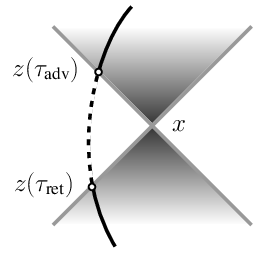}
  \caption{Singular term (dotted line).}
  \label{fig3}
\end{minipage}
\quad
\begin{minipage}[c]{0.45\textwidth}
    \includegraphics[width=3cm]{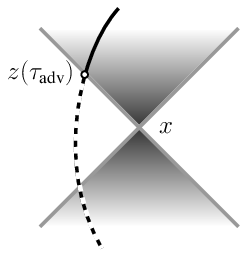}
  \caption{Radiative term (dotted line).}
  \label{fig4}
\end{minipage}

\end{figure}

In curved spacetime, Figs. \ref{fig1}-\ref{fig4}, at a given point $x$, the retarded term depends upon the particle history before the
retarded (proper) time $\tau_{\rm ret}$, whereas the advanced term depends upon the particle history after the advanced (proper) time $\tau_{\rm adv}$.
Figures (\ref{fig1}-\ref{fig4}) show ($\tau$ is proper time, $x$ the evaluation point, and $z$ the particle position). 

The singular is based on the 
interval $\tau_{\rm ret}<\tau<\tau_{\rm adv}$. 
The subtraction $Ret - S\!ing$ in curved space determines again a singularity-free quantity, but still depending upon the contributions from inside of the light cone, past and future. The inclusion of an additional, purposely built, function $H$ deletes such dependence   

\begin{align} 
Rad = Ret - S\!ing = Ret - \frac{1}{2}[Ret + Adv - H] = \frac{1}{2}[Ret - Adv + H]~~,
\end{align}      
where the {\it ad hoc} function $H$ is defined to agree with the advanced term when the particle position is in the future of the evaluation point, thereby cancelling the $S\!ing$ term   (the $Ret$ term is zero, for $\tau>\tau_{\rm adv}$). Finally, we have 

\begin{align}
Rad_{~\tau>\tau_{\rm adv}} = 0~~.
\end{align}
 
Notably, $H$ is defined to agree with the retarded term when the particle position is in the past of the evaluation point, also cancelling the 
$S\!ing$ term  (the $Adv$ term is zero, for $\tau<\tau_{\rm ret}$). Finally, we have 
\begin{align}
Rad_{~\tau <\tau_{\rm ret}} = Ret ~~.
\end{align} 

Furthermore, $H$ differs from zero at the intermediate values of the world-line outside the light-cone, between $\tau_{\rm ret}$ and 
$\tau_{\rm adv}$. The radiative component includes the state of motion at all times prior to the advanced time, and it is not a representation of the physical field but rather of an effective field. Remarkably, $H$ tends to zero when the evaluation point is at the particle position. 

The perturbation $h_{\mu\nu}$ is the difference between 
the full metric of the perturbed spacetime,
and the background. 
The Deteweiler-Whiting (DeWh) approach eùphasises that the motion occurs on a geodesic of the metric $g_{\mu\nu} + h_{\mu\nu}^{\rm R} $ where 
$h_{\mu\nu}^{\rm R} $ is the radiative part of the perturbation $h_{\mu\nu}$, and it implies two notable features: the regularity of the radiative field and the
avoidance of any non-causal behaviour. 

In the MiSaTaQuWa description \cite{Mino-Sasaki-Tanaka-1997,quwa97}, the perturbation $h$ is split into instantaneous (or direct) and tail (or indirect) parts. The former
is the radiation that goes to infinity along the light-cone, the latter accounts for the fact that gravitational perturbations can scatter off the background curvature and interact with the particle at later times.

For the quantitative definition of $h_{\mu\nu}^{\rm R} $, and for the relation with $h_{\mu\nu}^{\rm tail}$, see \cite{Poisson-Pound-Vega-2011}. Avoiding the description of the many technical difficulties, wecan derive an expression that is somewhat the difference between the geodesic in the background and the geodesic in the full (background + perturbation) metric

\beq
\frac{D^2 \Delta z^\alpha}{d\tau^2}=
\underbrace{- {R_{\mu\beta\nu}}^\alpha u^\mu \Delta z^\beta u^\nu}_{Background~metric~geodesic~deviation} 
\underbrace{- \frac{1}{2}
(g^{\alpha\beta} + u^\alpha u^\beta) 
(2h_{\mu\beta ;\nu}^{\rm tail} - h_{\mu\nu ;\beta}^{\rm tail} ) u^\mu u^\nu}_{Self-acceleration}~~. 
\label{gweq}
\eeq

Stemmed from geodesic principles, a geodesic deviation equation is thus obtained by subtracting the background from the perturbed motion, Eq. (\ref{gweq}). This equation has appeared first in \cite{Gralla-Wald-2008,Gralla-Wald-2011}, with a different derivation. The first right-hand side term depends on the background metric, while the second right-hand term depends upon the perturbations and it is the non-trivial 
self-acceleration term. The latter, multiplied by $m$, provides the known MiSaTaQuWa equation description \cite{Mino-Sasaki-Tanaka-1997,quwa97}.

The interpretation of Eq.(\ref{gweq}) leads to consider the self-acceleration term (second right hand-side term) causing a displacement in the trajectory represented by the geodesic deviation in the background metric (first tiht hand-side term). 

\subsection{The self-force for a massless particle}\label{sec:SFmassive}

The idea is to describe the effect of the self-force and of the geodesic deviation on a photon as an observable frequency shift. To this end, we replace $F^\alpha / m$ with the expression $c^2 k^\alpha$, keeping the same dimensions of an acceleration. A force on an object is given by the change of momentum over time. For massless particles, a time varying momentum implies a change in direction 
(scattering) and/or a change in magnitude (frequency shift and/or change of velocity). In GR, particles propagate along geodesics, and massless particles propagate along null geodesics. If we consider that the photon remains on the same geodesic, instead of stepping into a new one \cite{Gralla-Wald-2008,Gralla-Wald-2011}, we avoid the issue of blurred images, that concerns observational astronomy.
Therefore, the influence of the self-force could be interpreted as a shift in the photon frequency. Here, $k^\alpha = \left(\displaystyle{\frac{\omega}{c}}, \vec{k} \right)$ is the wave-vector. 

Assuming that the latter is orthogonal to the 4-velocity $s^\alpha$, we apply the projection operator $\delta^\alpha_\beta + s^\alpha s_\beta$ to the geodesic equation to obtain the expression for the frequency shift

\begin{equation}
    k^\alpha=-\frac{1}{2c^2}\left(g^{\alpha\beta}+s^\alpha s^\beta\right)\left(2\nabla_\delta h_{\beta \gamma}- \nabla_\gamma h_{\beta\delta}\right)s^\gamma s^\delta.  
\end{equation} \label{eq:final}

\section{Conclusion}\label{sec:Conclusion}

In this work, we have developed a new formalism for deriving the Regge-Wheeler-Zerilli equations for massless particles, through the generalisation of the Aichelburg-Sexl Stress-Energy-Momentum tensor. We include angular components which allows our geodesics to describe more than just radial falls. 
After deriving the source equation for the generalised metric, we offer arguments to consider the effects of gravitational self-force on massless particles as a mechanism for frequency shifts. The next logical step would be to solve the RWZ equations using numerical methods in order to recover the perturbation tensor and estimate the scale of the frequency shifts at hand. The concept of the self-force on photons could be studied in cosmology using the Friedmann-Lema\^itre-Robertson-Walker metric.

\bibliographystyle{unsrt}
\bibliography{references_spallicci_250703}
\end{document}